\documentclass[12pt]{article}

\usepackage{graphicx,subfigure,tabularx,booktabs}
\usepackage{float}
\usepackage{soul}
\usepackage[dvipsnames]{xcolor}
\usepackage{upgreek}
\usepackage{times}
\newcommand{\um}{$\upmu$m }

\usepackage{lineno}
\usepackage{changes}
\definechangesauthor[name=Adrian, color=red]{AL}

  \usepackage[
    backend=biber,
   style = nature,
    doi = false,
    url = false,
    isbn = false,
  ]{biblatex}
  
\let\cite=\supercite

\usepackage{xr}
\makeatletter
\newcommand*{\addFileDependency}[1]{
  \typeout{(#1)}
  \@addtofilelist{#1}
  \IfFileExists{#1}{}{\typeout{No file #1.}}
}
\makeatother

\newcommand*{\myexternaldocument}[1]{
    \externaldocument{#1}
    \addFileDependency{#1.tex}
    \addFileDependency{#1.aux}
}

\myexternaldocument{Supplementary_Information}

\title{
High Absorptivity Nanotextured Powders for Additive Manufacturing}

\author
{Ottman A. Tertuliano,$^{1,2\ast\dagger}$ Philip J. DePond,$^{2,3\dagger}$   Andrew C. Lee$^4$, \\Jiho Hong,$^4$ David Doan$^2$,Luc Capaldi$^1$, Mark Brongersma$^4$,\\ X. Wendy Gu,$^2$ Manyalibo J. Matthews,$^3$ Wei Cai, $^2$ Adrian J. Lew$^{2\ast}$\\
\\
\normalsize{$^{1}$Mechanical Engineering and Applied Mechanics, University of Pennsylvania}\\
\normalsize{220 S. 33rd St., Philadelphia, 19104, PA, U.S}\\
\normalsize{$^{2}$Mechanical Engineering, Stanford University}\\
\normalsize{452 Escondido Mall, Stanford, 94305, CA, U.S}\\
\normalsize{$^{3}$Materials Science Division, Lawrence Livermore National Laboratory}\\
\normalsize{7000 East Ave, Livermore, 94550, CA, U.S.}\\
\normalsize{$^{4}$Materials Science and Engineering, Stanford University}\\
\normalsize{496 Lomita Mall Suite 102, Stanford, 94305, CA, U.S.}\\
\normalsize{$^\dagger$equal contributions}\\
\normalsize{$^\ast$oat@seas.upenn.edu, $^\ast$lewa@stanford.edu}
}

\date{\vskip -12pt}

\begin{document} 

\baselineskip24pt 

    \maketitle 
\clearpage
\vskip -20pt
\textbf{Abstract: The widespread application of metal additive manufacturing (AM) is limited by the ability to control the complex interactions between the energy source and the feedstock material. Here we develop a generalizable process to introduce nanoscale grooves to the surface of metal powders which increases the powder absorptivity by up to 70\% during laser powder bed fusion. Absorptivity enhancements in copper, copper-silver, and tungsten enables energy efficient manufacturing, with printing of pure copper at relative densities up to 92\% using laser energy densities as low as 82 J/mm$^3$. Simulations show the enhanced powder absorptivity results from plasmon-enabled light concentration in nanoscale grooves combined with multiple scattering events. The approach taken here demonstrates a general method to enhance the absorptivity and printability of reflective and refractory metal powders by changing the surface morphology of the feedstock without altering its composition.}

\clearpage

Metal additive manufacturing (AM) has wide spread potential in the health care, aerospace, automotive and energy industries~\cite{DebRoy2018}. To date, the application of metal AM is constrained by a small library of weldable materials which can be reliably printed without extensive post-processing, e.g.,  stainless steels, AlSi10Mg, some Ni superalloys and Ti alloys~\cite{Liu2019}. These materials are readily printable in commercial laser powder bed fusion (LPBF) systems which typically use a near-infrared (1060-1080 nm wavelength) laser to scan over a layer of metal powder to induce melting and fuse the powders. Repeating this process layer-by-layer results in net shape 3D printed structures. However, free-form printing of non-weldable high reflectivity and refractory metals has been limited by the photothermal properties of the powder feedstocks\cite{Tran2019,Talignani2022AAlloys}. 
%

Unlike the aforementioned weldable metals, the low absorptivity (high reflectivity) in the near-infrared and high thermal diffusivity of copper, silver and their eutectic alloy make localizing heat during laser scanning challenging. Similar obstacles are encountered in LPBF of other high-reflectivity metals~\cite{Tran2019}, and illustrate a general need to control the \textit{in situ} photothermal properties and melting of powders during LPBF. Refractory metals like tungsten also have high thermal conductivities coupled with high melting points that pose further obstacles to additive manufacturing\cite{Morcos2022Review:Alloys,Talignani2022AAlloys}. Because of their high melting points, refractory metals experience large thermal gradients during LPBF that result in residual stresses and cracking after solidification\cite{Vrancken2020}. 
Fundamentally changing the interaction of powder feedstock with the energy source to improve absorptivity will expand the library of printable materials to include high reflectivity and refractory metals like copper, silver and tungsten.  

To date, approaches to enable newly printable metals have focused on changing their solidification and re-crystallization with  engineering solutions that employ additives. 
Adding nanoparticle inoculants to Al 7075 powder enabled printing of high-strength, crack-free structures~\cite{Martin2017b}. It has been reported that adding copper to Ti enabled forming ultrafine-grained alloys with tunable strength~\cite{Zhang2019}. A tailor-designed alloy, with functional  Fe19Ni5Ti, was developed to produce Damascus steel using laser powder bed fusion\cite{Kurnsteiner2020High-strengthManufacturing}. Tantalum and rare earth elements are added to tungsten in order to bind  traces of oxygen and mitigate cracking\cite{Xue2021SelectiveMicrocracks,Ikeshoji2018,Hu2020PureMechanism}. 



To specifically improve absorbed laser power, instrumentation is often modified to demonstrate printing of pure copper. High power infrared LPBF systems have been custom built to overcome the low absorptivity barrier, employing laser powers of 800 W and higher~\cite{Colopi2018,Colopi2019LimitsLaser,Ikeshoji2018}. These high power systems are reported to damage optical components, are not considered viable solutions~\cite{Jadhav2019}, and must also address the problem that the absorptivity of copper increases with temperature~\cite{Yilbas1991}. This increase makes controlling energy deposition into the copper melt pool difficult, an issue which escalates quickly at $\sim$ 1 kW laser powers~\cite{Colopi2019LimitsLaser}. Systems with high powered green lasers are being developed to take advantage of higher absorptivity of copper in the visible wavelengths\cite{Fu2022TheInnovation} but this can be prohibitively expensive for democratizing additive manufacturing. For managing detrimental temperature gradients, preheating is more commonly used, specifically with refractory metals\cite{Muller2019Additive1000C}, however these techniques may require substrate preheating up to 1000 $^\circ$C, and have proved most efficient with high energy density electron beam systems\cite{Dorow-Gerspach2021AdditiveMelting}. 

Beyond using high power systems to increase absorbed energy, adding nanoparticles to metal powders has been established as a promising approach to enable the printing of various metals~\cite{Martin2017b,Martin2020GrainAluminum,Jadhav2019,Jadhav2020,Jadhav2020HighlyPowder}. High purity copper and copper alloys have been printed by using nanoparticle-decorated copper surfaces~\cite{Jadhav2019,Jadhav2020}. These approaches demonstrated that the room temperature optical absorptivity of the decorated copper powder increases to $\sim$ 60\% in comparison to  $\sim$ 20\% for the uncoated copper. The use of additives in printing copper has also resulted in reduced electrical conductivity when chromium nitride was introduced to the surface copper powders~\cite{Jadhav2020} or solidification cracking when carbon nanoparticles are introduced~\cite{Jadhav2019}. The cracks observed in those studies originated at boundaries of segregated additives that did not melt, even at 0.1\% weight fraction. Graphene nanoflakes added to copper at similar weight fractions has resulted in increased absorptivity and relative density of printed parts~\cite{Tertuliano2021Nanoparticle-enhancedFusion}. While all the aforementioned processes have opened up new avenues of microstructural and photothermal control of powders, to date, no process exists which modifies powder feedstock without alloying or using additives and yet results in improved powder absorptivity, powder dynamics or print quality of difficult-to-weld high reflectivity and refractory metals. 

In this study, we developed an etching process for producing modified metal powder feedstock, and specifically, for improving absorptivity. The surface of conventional metal powders is modified using a wet chemical etching technique to produce nanoscale surface features. We demonstrate an increase in absorptivity, due to an increased localized absorption on the powder's nanoscale features using in-situ calorimetry experiments, electromagnetic (EM) simulations on single powder particle surfaces, and ray tracing simulations on a powder bed. Although the quality of printed parts is influenced by many factors beyond absorptivity, we demonstrate that these surface modified powders can enable printing of high purity copper and tungsten metal structures using lower power (100-500 W) laser-based metal 3D printing systems. The approach developed here enables printing of difficult-to-weld high reflectivity and refractory pure metals with the same energy requirements as commercially printed alloys. 

\section*{Results}

\subsection*{Etching Produces Nanoscale Surface Structures}

\begin{figure}[ht]
    \vskip -20pt
    \centering
    \includegraphics[width=0.7\textwidth]{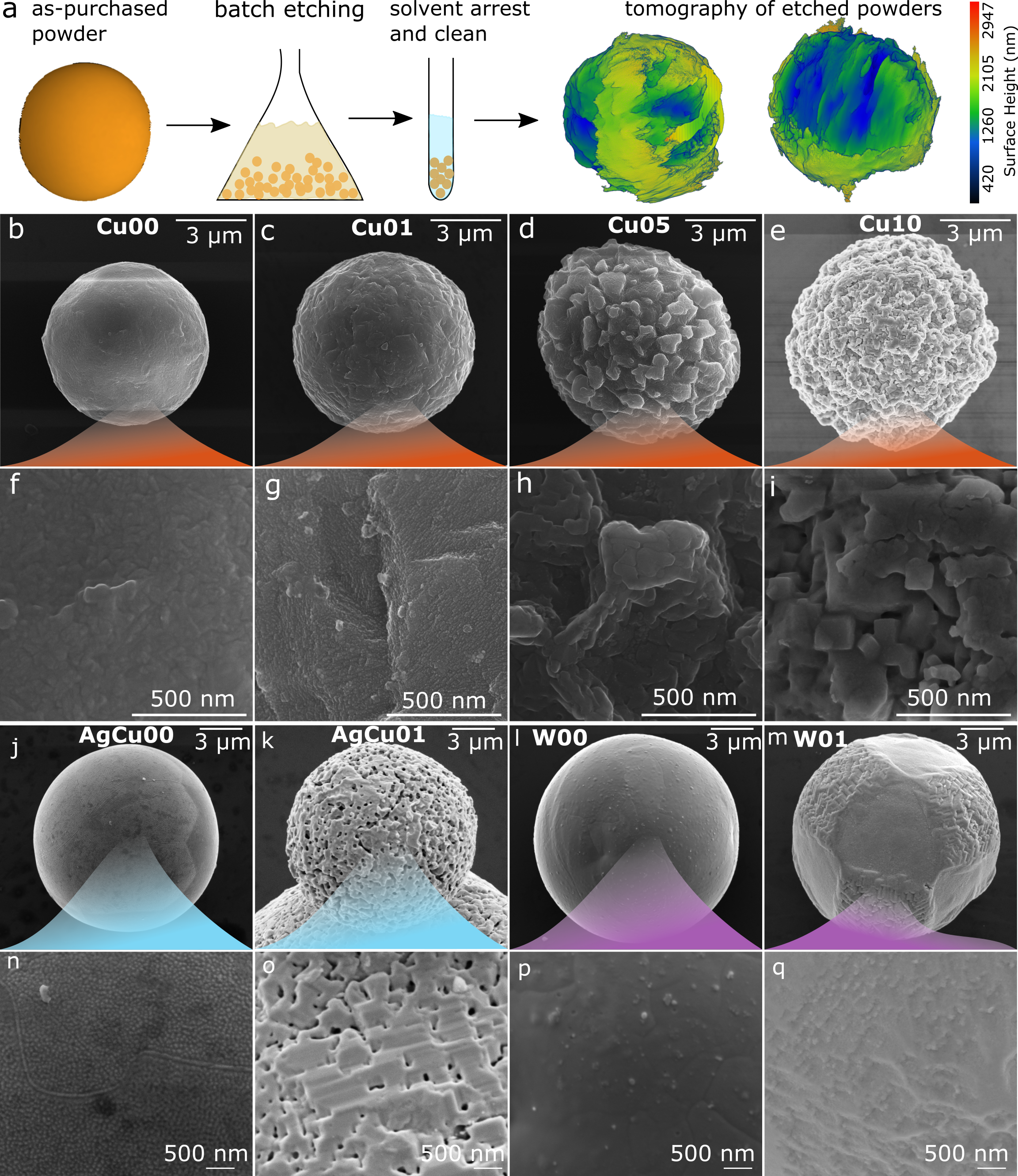}
    \caption{Surface of textured powders. a)Powder etching procedure with reconstructed 3D images from X-ray nanotomography of etched Cu powder showing surface topography. b) As-purchased (control) powder (Cu00). c) Powder particles etched for 1 h (Cu01), d) 5 h (Cu05), and e) 10 h (Cu10). f-i) High magnification images of powder surfaces showing progressively rougher features characterized by a change in feature size with etching time. Similar results are shown with: j) As-purchased AgCu powder and k) Etched AgCu powder, as well as l) As-purchased W powder and etched W powder. n-q) High magnification images of powder surfaces from (j-m)}.
    \label{fig:etched_powders}
\end{figure}

We produced nanotextured copper, copper-silver and tungsten powders via a batch solution process. We developed nanotextured copper powders by etching as-purchased (LPW, 99.95 \% purity) and as-fabricated (LLNL 99.99 \% purity) copper powders using a solution of FeCl$_3$, HCl, and ethanol (Methods, Fig.~\ref{figS:etching_procedure}). The results here are primarily reported for the LPW copper powder with consistent results for the LLNL powder reported in the Supplementary Information. The as-purchased LPW powder was processed for etch times ranging from 1 to 10 h (Fig.~\ref{fig:etched_powders}a) to produce powders with varying surface structures. As shown in the scanning electron microscopy (SEM) image in Fig.~\ref{fig:etched_powders}b, the surface of the as-purchased powder initially appeared smooth. After 1 h of etching (Fig.~\ref{fig:etched_powders}c), the powder began to exhibit uniform roughness on the surface. Figure~\ref{fig:etched_powders}d shows that etching for 5 h results in surface structures with substantially etched grain boundaries in addition to the uniformly etched grain surfaces observed after 1 h. This is likely due to a high etching selectivity of grain boundaries. Figure~\ref{fig:etched_powders}e reveals that after etching for 10 h, grain boundaries became highly visible, as well as the presence of cubic structures with characteristic dimensions on the order of 100 nm on the surface of the powders. Figures~\ref{fig:etched_powders}f-i show magnified regions of the surface of the powder particles shown in Fig.~\ref{fig:etched_powders}b-e, respectively. These magnified images show progressively rougher surfaces characterized by an increase in feature size with etching time up to 5 h. 
We hereon refer to the four powder types based on the etching time as follows: Cu00, Cu01, Cu05, and Cu10, where CuX indicates X hours of etching. A similar etching procedure and nomenclature is used for AgCu (Fig.~\ref{fig:etched_powders}j,k and n,o) and W powders (Fig.~\ref{fig:etched_powders}l,m and p,q, see Materials and Methods). These results demonstrate the generalizability  of the powder etching procedure to high reflectivity and refractory metals.

We quantified the etching by calculating the effective volumetric etch rate using the Cu05 nanotomography results show in Fig.~\ref{fig:etched_powders}a. The results show tomography of a Cu05 powder particle from two different points of view. By comparing the volume of the etched powder particle to that of its convex hull  (Fig.~\ref{figs:convex_hull}), we estimated an effective volumetric etch rate of 11 $\upmu$m$^3$/h. 
%
%
For the specific powder particle measured, we calculated an effective surface depth etch rate of about 71 nm/h for up to 5 h of etching. This would be considered a lower bound as we do not account for  uniform etching. Further details are provided in Supplementary Information.



\subsection*{Nanotextured Surfaces Increase Powder Absorptivity}
\begin{figure}[ht] 
    \centering
    \includegraphics[width=0.85\textwidth]{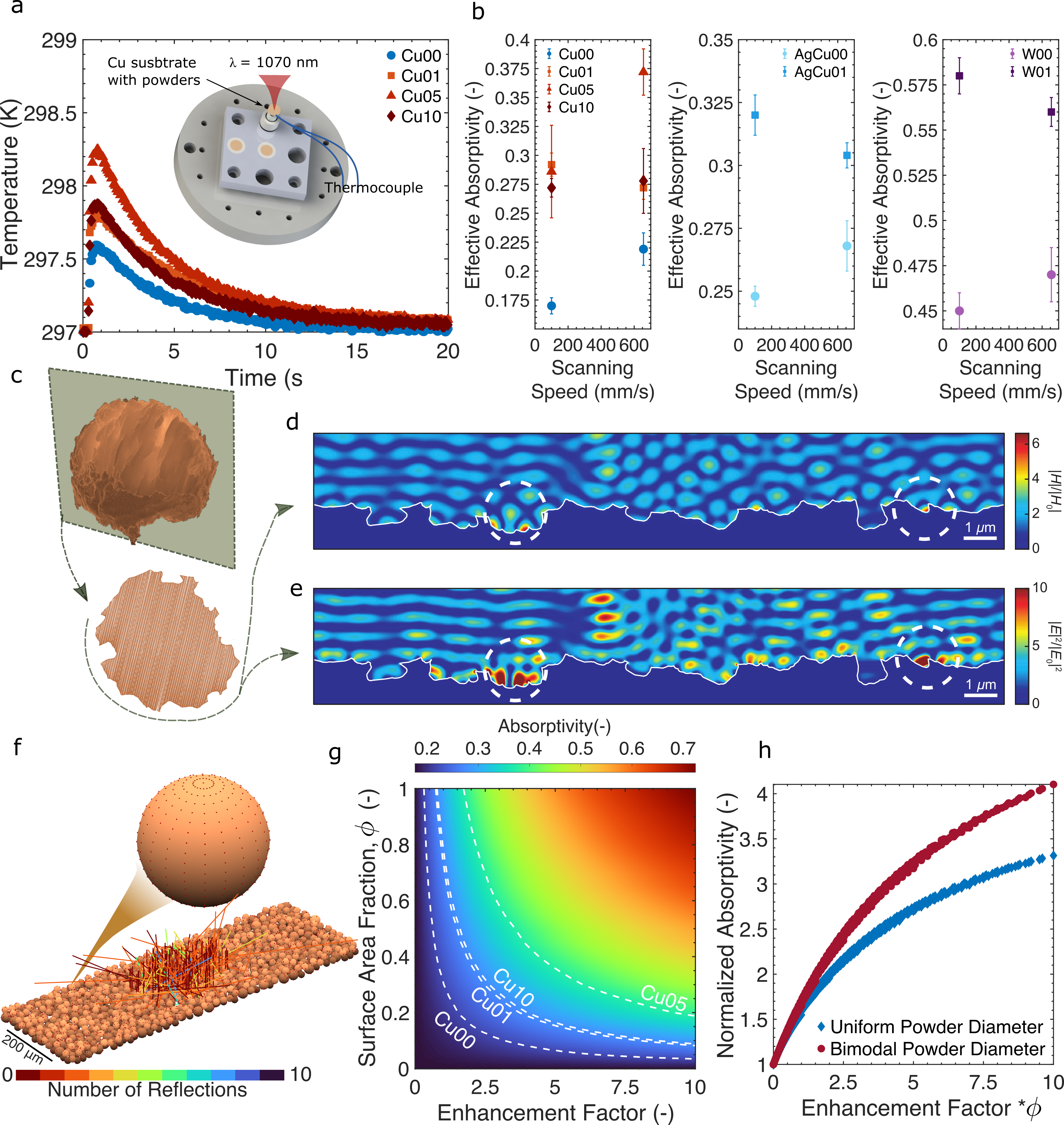}
    \caption{Absorptivity enhancement in etched powder. a) Representative time vs. temperature data from calorimetry experiments performed at 175 W and 656 mm/s. Data shows an increase in peak substrate temperature in etched powder relative to as-purchased powders. Inset shows \textit{in situ} calorimetry experimental setup. b) Effective absorptivity in as-purchased (blue) and etched copper powders at a laser power of 175 W and two speeds, 100 and 656 mm/s, showing increase in effective absorptivity of etched powder relative to as-purchased copper powders at both scan speeds. Similar results shown for AgCu and W as-purchased and etched powders. c) Sample particle cross-section used for EM simulations. d) Normalized magnetic field and e) Electric field intensities showing localized fields in surface grooves. f) Representative simulation domain for ray tracing calculations. The colors of the rays represent the number of reflections of each ray, where incident rays are assigned a value of 0. The red spots on the powder particle surfaces indicate regions of enhanced absorptivity, covering a surface area fraction  $\phi$. g) Absorptivity map from ray tracing simulations of bimodally distributed Cu powder diameter. Dashed lines are iso-absorptivity contours corresponding to the measured absorptivity of Cu powders at 656 mm/s. h) Simulation results showing that  absorptivity improves faster in bimodally distributed powders than in uniformly distributed powder. The absorptivity is normalized by the respective values at $\phi=0$ for each powder distributions.}
  
    \label{fig:absorptivity}
\end{figure}

Calorimetry experiments conducted at a laser power of 175 W and a $1/e^2$ beam diameter of 60 \um revealed an improved absorptivity in the nanotextured powders over the as-purchased powders (Fig.~\ref{fig:absorptivity}a,b), regardless of material. We measured the effective absorptivity, $A_{\rm eff}$, of the Cu00 powder to be 0.172 at scanning speed 100 mm/s and 0.219 at a scanning speed of 656 mm/s.
%
At the slower speed of 100 mm/s, the etched powders exhibit improved absorptivities of 0.292, 0.286 and 0.272, for Cu01, Cu05, and Cu10, respectively. At the faster scanning speed of 656 mm/s we measured the corresponding absorptivities of 0.272, 0.372 and 0.278 for the etched powders. 
At both scanning speeds, the nanotextured powders exhibited an absorptivity enhancement factor (absorptivity normalized by that of as-purchased powder absorptivity) of up to 1.7. The Cu05 powder at the faster scan speed provides the highest absorptivity overall. These results are summarized in Table~\ref{table:absorptivity}. We demonstrated similar improvement in absorptivity (enhancement factor of 1.5) using the higher purity LLNL copper powder (Fig.~\ref{figS:LLNL_etch_calorimetry}). AgCu and W demonstrate absorptivity enhancement factors up to 1.3, with W increasing from 0.45 to 0.58 (Fig.~\ref{fig:absorptivity}b). 


To gain a better understanding of how nanotexturing enhanced absorptivity on an individual powder particle, we performed EM wave simulations of light absorption on a textured planar surface (see Methods). The surface profile was extracted from the cross-section of the reconstructed Cu powder particle tomograph (Fig.~\ref{fig:absorptivity}c). The higher total absorptivity of the etched powders results from enhanced light-matter interaction on the textured Cu surfaces. 
For an incident plane wave with transverse-magnetic polarization (i.e., electric field in the in-plane directions), the simulated field distributions (Fig.~\ref{fig:absorptivity}d,e) show that certain grooves in the surface (dashed circles) provide strong near-field intensities and boost local absorption\cite{Qin2017}. This result suggests only a fraction of the powder area leads to the increased absorption of the textured surface compared to the flat one. This enhanced absorption on a single etched particle occurs over broad range of incident angles ($\pm60^o$); it results in an average absorption enhancement factor, i.e., the absorption of nanostructure normalized by that of a flat copper substrate, of 1.8 (Fig.~\ref{figS:enhancement_vs_angle}). 

To understand how the enhanced absorptivity of a single etched particle affects that of the powder bed, we performed ray tracing simulations on unmelted powder. Informed by the EM simulations, we treat the nanotextured powders as having a fraction of their surfaces, $\phi$, with a higher local absorptivity, as shown in Fig.~\ref{fig:absorptivity}f. We simulated powder beds with uniform and bimodal (20 and 40 \um) particle diameter distributions sizes with a 30 \um mean diameter. In each powder bed distribution we explored the full range of surface area fraction ($\phi\in[0 \;, 1])$ with local absorption enhancements factors ranging from 1 to 10. 
To check consistency, we calculated the effective absorptivity of untreated Cu00 powders ($\phi = 0$ and an enhancement factor of 1). The predicted absorptivity is 0.240 and 0.176 in the uniform and bimodal diameter distributions, respectively, showing agreement with the measured value of 0.219 (Fig.~\ref{fig:absorptivity}b). Details of the simulation are provided in the Methods.

Figure~\ref{fig:absorptivity}g shows an absorptivity map from simulation results that illustrate possible combinations of $\phi$ and local enhancement factors for the powders in this work.  The iso-absorptivity contours in the same map correspond to the experimental results for 656 mm/s. Based on the EM simulations in Fig.~\ref{fig:absorptivity}d,e, we expect the etched powders to have low values of $\phi$, meaning higher local enhancement factors to enable the same absorptivity on the contour lines of Fig.~\ref{fig:absorptivity}g. The iso-absorptivity contours suggest that the absorptivity is largely determined by and is an increasing function of the product of the local enhancement factor and the surface area fraction. 
The absorptivity increases faster with this product in bimodally distributed powder than in uniformly distributed powder with the same average diameter (Fig~\ref{fig:absorptivity}h). These results demonstrate that the manner in which modulating single particle absorptivity influences powder bed absorption is coupled to the powder bed particle size distribution, presenting a multiscale enhancement mechanism.

\subsection*{Nanotextured Powders Show Improved Printing at Low Powers}

\begin{figure}[ht]
    \centering
    \includegraphics[width=1\textwidth]{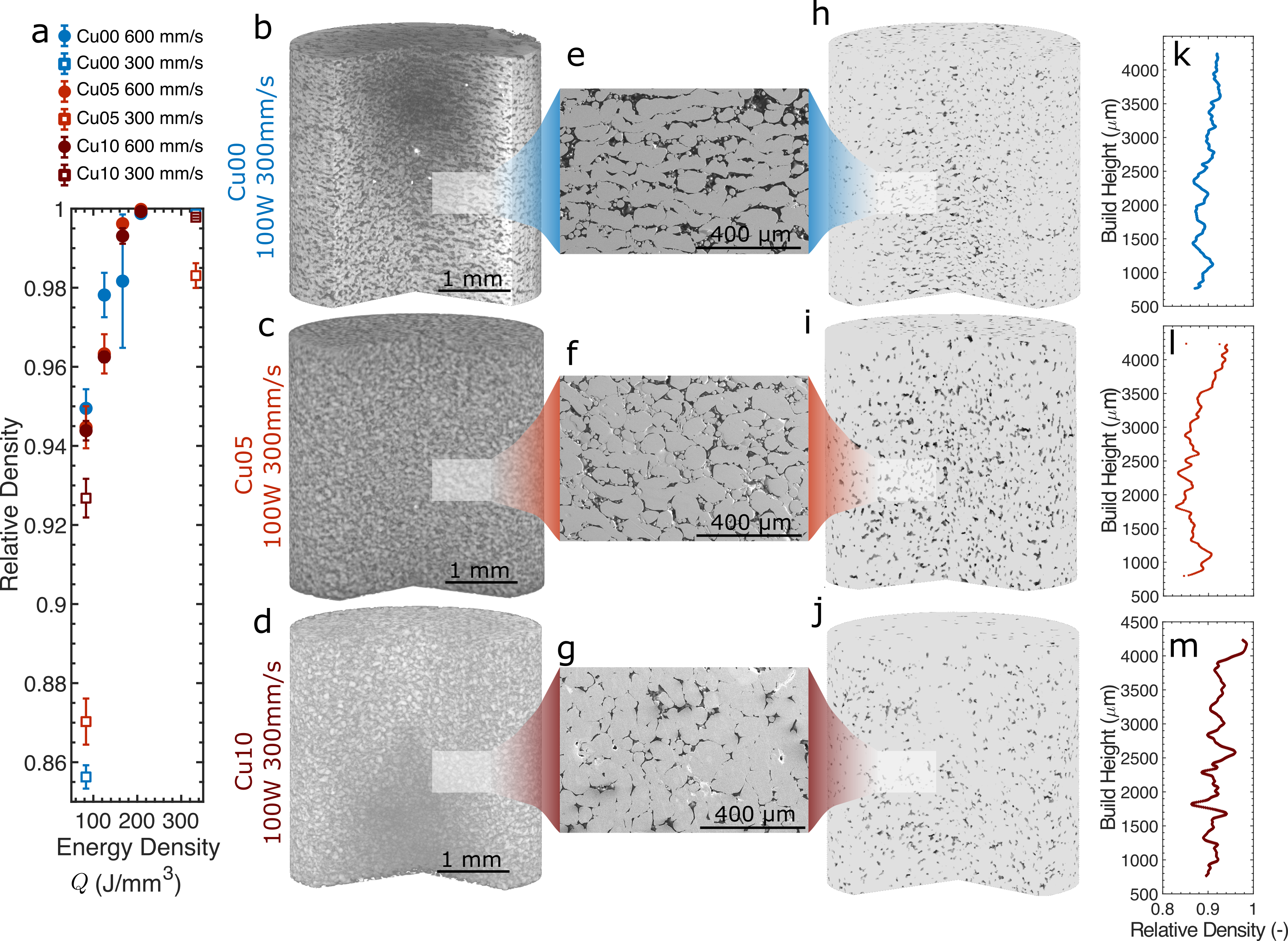}  
    \caption{Print quality. a) Relative density as a function of volumetric energy density, $Q$.
    b-d) X-ray computed tomography results of a cylindrical volume extracted from the center of 6 mm-cylinder builds for $Q$=83 J/mm$^3$ . e-g) SEM cross-sections of builds in b-d, respectively. h-j) Binarized tomography results using a watershed scheme. k-m) Relative density as a function of build height averaged over 55 \um increments.}
    \label{fig:relative_density}
\end{figure}

To assess the viability of high absorptivity nanotextured powders for LPBF, we printed cylindrical structures. We quantified the relative density (i.e., the solid volume fraction) as a function of the laser scanning parameters consolidated as the volumetric energy density, $Q = \frac{P}{htv}$ (see Table~\ref{table:relative density}), where $P$ and $v$ are the laser power and speed, and $h$ and $t$ are the hatch spacing and powder layer thickness. At lowest values of $Q$ (83 J/mm$^3$, calculated for the scanning condition of $P$=100 W and $v$=300 mm/s, and $P$=200 W and $v$=600 mm/s), the etched powders provided an improvement in relative density over the as-purchased powders (Fig.~\ref{fig:relative_density}a). At 100 W and 300 mm/s the print using the Cu10 powders has density of 0.926 with a measurement error of $\pm$ 0.004 (Fig.~\ref{fig:relative_density}a,g), an improvement over the 0.856 $\pm$ 0.003 observed in the Cu00 powder (Fig.~\ref{fig:relative_density}a,e). The Cu05 powder demonstrates similar relative density of 0.870 $\pm$ 0.005 (Fig.~\ref{fig:relative_density}a,f) as the  as-purchased powder. Under these conditions, porosity stems primarily from lack-of-fusion defects (Fig.~\ref{fig:relative_density}e-f). At the same $Q$, but with twice the power and speed (200 W and 600 mm/s), the different powders produced similar quality builds with relative densities. As $Q$ was increased beyond 200 J/mm$^3$, the relative density converged to around 0.98-0.99 for all prints regardless of powder treatment. It has been shown that laser-powder interactions become less relevant at high powers (greater than 200 W in stainless steel) because the beam simply resides on top of the melt pool rather than interacting with the powder~\cite{Khairallah2020ControllingPrinting}. Our results are consistent with that observation as the improvement in print quality is most beneficial at lower energy inputs. 

Relative density measurements from tomography of the 100 W printing condition samples, shown in Fig.~\ref{fig:relative_density}b-d, are consistent with those acquired from the SEM images in Fig.~\ref{fig:relative_density}a. The tomography reveals that at  low powers, the nanotextured powders may exhibit more fluctuations in relative density as a function of build height (Fig.~\ref{fig:relative_density}l,m), compared to  that observed in the as-purchased powder. This  suggests that at lower powers there may be different laser-powder interactions and melt and powder dynamics in these higher absorptivity nanotextured powders relative to the as-purchased powders\cite{Khairallah2020ControllingPrinting}. At high powers of 400 W, these differences are less relevant as we observe high relative density prints (Fig.~\ref{fig:relative_density}a, Fig.~\ref{figS:prints_400W},~\ref{figS:relative_density}).

\section*{Discussion}


\subsection*{Evolution of Surface Nanotexure}

The surface morphology results in copper suggest that the powder undergoes three main stages during the etching process: uniform etching, grain boundary etching and redeposition. We can better understand the powder evolution through these stages as the etching of Cu in FeCl$_3$ solutions occurs with two reactions. The FeCl$_3$ strips Cu from the powder surface to create CuCl$_2$ in solution. This CuCl$_2$ in solution further acts as a secondary etchant by creating complexes with Cu from the powder surface to create 2CuCl. In a three-stage etching process, uniform etching occurs first in the 1 h time scale due to dissociation of Cu from the surface by forming complexes with Cl$^{-}$ ions in the solution.  Second, in the 5 h time scale, grain boundary etching becomes evident, as the etching selectivity of grain boundaries is higher than in the bulk \cite{Young1961EtchCopper}. This is demonstrated further in supplementary images of Cu05 powder (Fig.~\ref{figS:etched_grain_boundaries}). 

The third step occurs when further processing of the powders between the 5 and 10 h results in redeposition of cubic nanocrystals on the surface of the powder particles, without altering the feedstock composition. Energy dispersive spectroscopy in an SEM has indicated that these nanocrystals are primarily copper (Fig.~\ref{figS:eds}). Copper in the etching solution may exist as CuCl$_2$ or 2CuCl and the nucleation of the observed cubic structures on the powder surface may be energetically favorable in the concentrations achieved before 10 h of etching. The Cu nucleates in highly faceted cubic morphology with orthogonal faces, indicating preferential growth of \{100\} crystal planes, as observed in Fig.~\ref{fig:etched_powders}h. For the observed cubic structures to emerge, two conditions should have been met in the etching solution: 1) the solubility limit of 2CuCl in the diluted HCl ($\sim$ 5g/100 mL) should have been reached and 2) there should be a preferred dissociation of Cu from Cl such that Cu crystals can nucleate and grow on Cu powder surfaces. Using the volumetric etch rate of 11 $\upmu$m$^3$/h from the single particle nano-tomography, we estimate that 100 g of powder with a mean particle diameter of 30 \um dissolves at about 86 mg/h. This results in about 0.86 g of Cu available to create 1.3 g of CuCl in the 100 ml of etching solution in 10 h of etching. This amount of CuCl is consistent with the solubility limit in HCl, within an order magnitude, and supports redeposition as a mechanism of creating the cubic structure observed on the Cu10 powders (Fig.~\ref{fig:etched_powders}i). Thus, etching to produce surface features characteristic of the powder grain structure requires either a shorter time (e.g., 5 h in this work) or dilute enough solution in active species to prevent re-nucleation. 

\subsection*{Self-evolved surface nanotexture modifies \textit{in situ} laser-powder interactions}


The measured absorptivity of the all nanotextured powders increases relative to that of all the as-purchased powders under all etching conditions. Previous efforts aimed at improving absorptivity of powders in LBPF involve reducing average powder size\cite{Qu2021High-precisionCopper,Silbernagel2019}, alloying \cite{Jadhav2020HighlyPowder}, or using additives \cite{Tertuliano2021Nanoparticle-enhancedFusion,Jadhav2019}. These methods have inherent limitations in powder handling and chemical composition. Our results demonstrate the ability to alter absorptivity of metal powder feedstock, using surface topography, without changing composition. The concept of using nanoscale topography to modify absorptivity is observed in nature and is a well-explored principle in flat optical devices
\cite{Vukusic2020-Nature,Shi2015KeepingAnts,Shimada2016WhatNanostructures,BurresiBright-WhiteLight,Teperik2008OmnidirectionalSurfaces}. The approaches used to modify absorptivity of flat surfaces employ carefully designed 2D, lithographically generated patterns on substrates to achieve a desired absorptivity or light-matter interaction \cite{Polman2012PhotonicPhotovoltaics,Teperik2008OmnidirectionalSurfaces}. Lithographic patterning to achieve such surfaces is not easily translatable and scalable to spherical powder surfaces, especially for metal AM. Here we produce self-evolved nanostructures by leveraging the anisotropic solution-based etching of metallic grains. This strategy enables high-throughput production of powders with improved absorptivity for metal AM. 
 
While all the self-evolved nanotextured powders showed improvements in absorptivity, the Cu05 powder provided the largest absorptivity enhancement under the explored laser scanning conditions. The beam conditions of 175 W and 656 mm/s resulted in a low $Q$ value of 67 J/mm$^3$ and the highest $A_{\rm eff}$ measured in the Cu05 powder. At such low energy densities applied to copper powder, we expect the laser to interact with fully intact or partially sintered powder rather than a fully-formed melt pool~\cite{Trapp2017}. This lower energy density beam conditions result in the absorptivity being measured in the conduction regime in copper \cite{Tertuliano2022Nanoparticle-enhancedFusion}, where the absorptivity enhancement from multiple reflects in a keyholing regime does not contribute. Modification of powder particle surface geometry is expected to have a larger effect on absorptivity at low energy densities relative to scanning at high power and slow speeds. Based on the feature sizes on the powder particle surfaces, we expect that nanoscale grooves provide regions of high absorptivity~\cite{Teperik2008OmnidirectionalSurfaces,Sndergaard2012PlasmonicGrooves}.

 

For nanoscale groove features to efficiently improve absorptivity, the groove dimensions are critical to induce optical or plasmonic resonance and localized heating at the laser wavelength\cite{Baffou2013}. The EM simulations on a full surface contour of a single powder particle provide an absorption enhancement factor of 1.8 (Fig.~\ref{figS:enhancement_vs_angle}) that is consistent with the measured value of 1.7. The strong electromagnetic fields at the surface are mainly attributed to plasmonic resonances supported by individual grooves \cite{Sndergaard2012PlasmonicGrooves,Kravets2008PlasmonicCoatings}. Understanding the drop in absorption enhancement from Cu05 to Cu10 requires understanding field localization in relation to local groove dimensions on particle surfaces. 





We chose representative grooves in the extracted surface profile and performed the EM simulation of each individual groove on a flat surface. The simulated absorption spectrum displays an optical resonance around the laser wavelength used in the experiment (Fig.~\ref{figS:wide_grooves_resonance}). On resonance, the simulated field distributions show large field magnitudes inside the groove and along the air-Cu interface, similar to those of the entire extracted surface profile. Due to its relatively low optical quality-factor (Fig.~\ref{figS:wide_grooves_resonance},c), the plasmonic resonances supported by the grooves also display a broad angular response~\cite{Teperik2008OmnidirectionalSurfaces}. Since the general behavior of such resonances depends on the dimensions of the resonator, simple rectangular grooves can provide similar mode profiles and spectral responses to those of experimentally extracted ones and be used to study the trend of the change in absorption as a function of the etching time. Through parametric sweeps, we observed that the absorption is increased for taller grooves, especially with subwavelength widths (Fig.~\ref{figS:deep_groove_resonance}). The incident light can funnel into the narrow grooves due to their strong near-field magnitude and be absorbed while propagating along the gap between the sidewalls of the grooves. The in-coupled light can also be reflected at the top and bottom of the grooves and this gives rise to plasmonic resonances that are mainly governed by the groove height. The Cu10 powder has shallower surface grooves relative to Cu05 (Fig~\ref{fig:etched_powders}g,h), due to redeposition of Cu, which can explain the measured drop in absorptivity.

The textured surface also has grooves with large widths, particularly Cu05. Such grooves can support higher-order plasmonic resonance in the lateral direction where the light propagates along the interface between their sidewalls, i.e., surface plasmon resonances (Fig.~\ref{figS:wide_grooves_resonance}). This also further contributes to the higher absorption of Cu05. The numerical analysis corroborates well with the experimental results that the absorption increases from Cu00 to Cu05 and decreases to Cu10.

Ray tracing simulations suggest that the electric field localization mechanism of absorption enhancement at the nanoscale on particle surfaces may contribute differently to the absorptivity of a powder bed. A parametric study changing $\phi$ and absorption enhancement factor to match experimentally measured powder bed absorptivities  showed the possible combinations of absorption enhancement factors and fractions of surface area contribution to absorption on each powder system (Fig.~\ref{fig:absorptivity}h). Specifically, ray tracing uniquely captures the coupling between particle absorption and powder bed absorption when the particle diameter distribution is altered. We find that we may improve the absorptivity of a non-uniformly (e.g., bimodal here) distributed powder bed at a faster rate than on uniformly distributed powder bed (Fig.~\ref{fig:absorptivity}i) by nanotexturing, i.e., improving single particle absorption~\cite{Boley2016,Boley2017}.



\subsection*{Printing with etched powders}
In the explored printing conditions, we observe the highest improvement in print quality using the Cu10 powders, as quantified by relative density measurements. The improvement is most pronounced at low energy densities, at which the surface structure of the powder is expected to play a larger role in light-matter interactions. However, the absorptivity measurements showed that Cu05 has a higher absorptivity than Cu10. The printing quality is influenced by other factors in addition to absorptivity. 
%
The higher absorptivity in Cu05 could result in recoil pressure induced expulsion of powder from the laser path. This could lead to increased denudation or back spatter, and may manifest as lack-of-fusion defects observed here~\cite{Khairallah2020ControllingPrinting}. 
 


\begin{figure}[ht]
    \centering
    \includegraphics[width =1\textwidth]{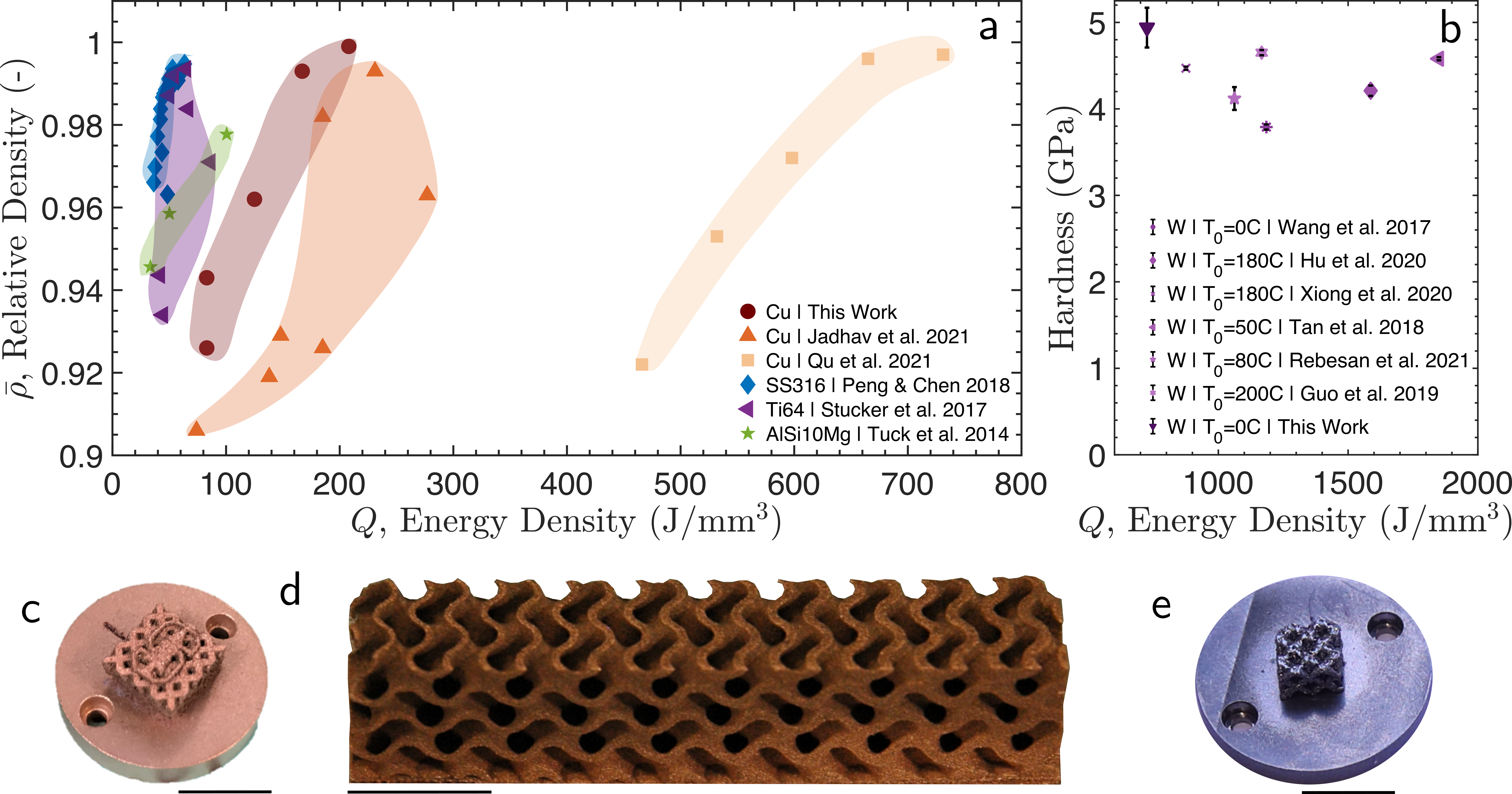}
    \caption{Print energy efficacy in LBPF. a) Readily printable materials such as SS316, Ti64 and Al alloys are printable with full relative densities ($\bar\rho \geq$ 0.99) using low energy densities ($Q\leq$ 80 J/mm$^3$). We demonstrate the ability to push the processing conditions of copper to lower energy densities, relative to previous works, using high absorptive Cu05.  Shaded areas show qualitative grouping of data. b) Indentation hardness of tungsten cylinder prints contextualized in energy density. The nanotextured W prints result in hardness of $\sim$ 5 GPa, a value similar to other measurements of additively manufacture pure W, but without the need to preheat powders up to 200 C (see legend). c and d) Printing of an octet lattice and a triply periodic minimal surface using Cu05 powders at 87 J/mm$^3$. e) Octet W structure printed using W01 at 725 J/mm$^3$. Scale bars are 10 mm.}
    \label{fig:print_efficiency}
\end{figure}

Given the new absorptivity behavior of the powders produced in this study, the optimal laser processing parameters for the etched powder should be different, (e.g. at lower energy density) from conventional Cu powders. It would be important moving forward to have dedicated studies on powder dynamics of modified powder feedstock as produced here. Despite these current limits in understanding fundamental powder dynamics, we demonstrate the utility of the high absorbing powders by printing full standing 50 mm long triply periodic minimal surfaces (Fig.~\ref{fig:print_efficiency}a insets, Fig.~\ref{figS:printing_demo}) at a power of 100 W and a scanning speed of 300 mm/s, a result not achievable using conventional, as-purchased copper powder. Figure~\ref{fig:print_efficiency}a generally illustrates the accessible processing conditions enabled by the nanotextured powder developed in this work, relative to other approaches. The nanotextured powder lowers the required energy densities for printing copper to levels similar to those for stainless steel and Ti alloys.

The print efficacy of the nanotextured powders was also contextualized by accessing the mechanical performance of printed tungsten via nanoindentation 
 in Fig.~\ref{fig:print_efficiency}b. Cylinders similar in size to Cu specimens printed with energy densities ranging from 500 to 1250 J/mm$^3$. We measure an indentation hardness of $\sim$ 5 GPa at an energy density of 725 J/mm$^3$, i.e, a hardness higher than that of other additively manufactured tungsten but achieved at lower energy densities\cite{Hu2020PureMechanism,Xiong2020SelectiveTungsten,Tan2018SelectiveProperties,Rebesan2021TungstenFusion,Guo2019SelectiveProperties,Wang2017DenseMelting}. Although the issue of cracking still remains to be solved in the printing of refractory metals, the higher absorptivity powders enable improved constitutive material properties with a fraction of the energy employed using other methods. 



 \section*{Summary and Outlook}

We demonstrated that the absorptivity of metal powder feedstock can be increased via self-evolving surface texture in an etching solution, without alloying or use of high absorbing nanoparticle additives. 
We attribute the increased absorptivity to the localization of incident light at nanoscale grooves on the powder surface, where  groove dimensions smaller or comparable to  the wavelength of the laser lead to resonances. The high absorptivity powders can enable printing starting at low energy densities (83 J/mm$^3$). These printing conditions have not been previously reported to print copper for the measured relative densities ($\geq$0.92). The powders developed here can be used for printing in moderately powered ($\sim$400 W) commercial laser powder bed fusion systems. These geometrically imperfect powders present a deviation from the idealized, smooth sphere morphology sought after in creating powder feedstock\cite{Vock2019PowdersReview}, yet present an improvement in photothermal efficiency and print quality in manufacturing. Our generalizable approach leverages feedstock surface imperfections to improve the laser–material interdependence without modifying the laser or the material composition.

\clearpage
\section{Code Availability}
Custom code written to process SEM images in Matlab to compute relative densities is available from corresponding authors without restriction, as well as all other processing code used in the work.
\printbibliography

@article{Martin2017b,
    title = {{3D printing of high-strength aluminium alloys}},
    year = {2017},
    journal = {Nature},
    author = {Martin, John H. and Yahata, Brennan D. and Hundley, Jacob M. and Mayer, Justin A. and Schaedler, Tobias A. and Pollock, Tresa M.},
    number = {7672},
    pages = {365--369},
    volume = {549},
    publisher = {Nature Publishing Group},
    url = {http://dx.doi.org/10.1038/nature23894},
    isbn = {0028-0836 1476-4687},
    doi = {10.1038/nature23894},
    issn = {14764687},
    pmid = {28933439}
}

@article{Tran2019,
    title = {{3D printing of highly pure copper}},
    year = {2019},
    journal = {Metals},
    author = {Tran, Thang Q. and Chinnappan, Amutha and Lee, Jeremy Kong Yoong and Loc, Nguyen Huu and Tran, Long T. and Wang, Gengjie and Kumar, Vishnu Vijay and Jayathilaka, W. A.D.M. and Ji, Dongxiao and Doddamani, Mrityunjay and Ramakrishna, Seeram},
    number = {7},
    pages = {12--20},
    volume = {9},
    doi = {10.3390/met9070756},
    issn = {20754701}
}

@article{DebRoy2018,
    title = {{Additive manufacturing of metallic components – Process, structure and properties}},
    year = {2018},
    journal = {Progress in Materials Science},
    author = {DebRoy, T. and Wei, H. L. and Zuback, J. S. and Mukherjee, T. and Elmer, J. W. and Milewski, J. O. and Beese, A. M. and Wilson-Heid, A. and De, A. and Zhang, W.},
    pages = {112--224},
    volume = {92},
    doi = {10.1016/j.pmatsci.2017.10.001},
    issn = {00796425}
}

@article{Liu2019,
    title = {{Additive manufacturing of Ti6Al4V alloy: A review}},
    year = {2019},
    journal = {Materials and Design},
    author = {Liu, Shunyu and Shin, Yung C.},
    pages = {107552},
    volume = {164},
    publisher = {The Authors},
    url = {https://doi.org/10.1016/j.matdes.2018.107552},
    doi = {10.1016/j.matdes.2018.107552},
    issn = {18734197}
}

@article{Zhang2019,
    title = {{Additive manufacturing of ultrafine-grained high-strength titanium alloys}},
    year = {2019},
    journal = {Nature},
    author = {Zhang, Duyao and Qiu, Dong and Gibson, Mark A. and Zheng, Yufeng and Fraser, Hamish L. and StJohn, David H. and Easton, Mark A.},
    number = {7785},
    pages = {91--95},
    volume = {576},
    publisher = {Springer US},
    url = {http://dx.doi.org/10.1038/s41586-019-1783-1},
    isbn = {4158601917831},
    doi = {10.1038/s41586-019-1783-1},
    issn = {14764687}
}

@article{Vrancken2020,
    title = {{Analysis of laser-induced microcracking in tungsten under additive manufacturing conditions: Experiment and simulation}},
    year = {2020},
    journal = {Acta Materialia},
    author = {Vrancken, Bey and Ganeriwala, Rishi K. and Matthews, Manyalibo J.},
    month = {8},
    volume = {194},
    doi = {10.1016/j.actamat.2020.04.060},
    issn = {13596454}
}

@article{Boley2017,
    title = {{Calculation of laser absorption by metal powders in additive manufacturing}},
    year = {2017},
    journal = {Additive Manufacturing Handbook: Product Development for the Defense Industry},
    author = {Boley, C. D. and Khairallah, Saad A. and Rubenchik, Alexander M.},
    number = {9},
    pages = {507--517},
    volume = {54},
    isbn = {9781482264098},
    doi = {10.1201/9781315119106}
}

@article{Silbernagel2019,
    title = {{Electrical resistivity of pure copper processed by medium-powered laser powder bed fusion additive manufacturing for use in electromagnetic applications}},
    year = {2019},
    journal = {Additive Manufacturing},
    author = {Silbernagel, Cassidy and Gargalis, Leonidas and Ashcroft, Ian and Hague, Richard and Galea, Michael and Dickens, Phill},
    number = {August},
    month = {10},
    pages = {100831},
    volume = {29},
    publisher = {Elsevier},
    url = {https://doi.org/10.1016/j.addma.2019.100831 https://linkinghub.elsevier.com/retrieve/pii/S2214860418309916},
    doi = {10.1016/j.addma.2019.100831},
    issn = {22148604}
}

@article{Trapp2017,
    title = {{In situ absorptivity measurements of metallic powders during laser powder-bed fusion additive manufacturing}},
    year = {2017},
    journal = {Applied Materials Today},
    author = {Trapp, Johannes and Rubenchik, Alexander M. and Guss, Gabe and Matthews, Manyalibo J.},
    pages = {341--349},
    volume = {9},
    publisher = {Elsevier Ltd},
    url = {http://dx.doi.org/10.1016/j.apmt.2017.08.006},
    doi = {10.1016/j.apmt.2017.08.006},
    issn = {23529407}
}

@article{Jadhav2019,
    title = {{Influence of Carbon Nanoparticle Addition (and Impurities) on Selective Laser Melting of Pure Copper}},
    year = {2019},
    journal = {Materials},
    author = {{Jadhav} and {Dadbakhsh} and {Vleugels} and {Hofkens} and {Puyvelde} and {Yang} and {Kruth} and {Humbeeck} and {Vanmeensel}},
    number = {15},
    pages = {2469},
    volume = {12},
    doi = {10.3390/ma12152469}
}

@article{Yilbas1991,
    title = {{Measurement of temperature-dependent reflectivity of Cu and Al in the range 30-1000 degrees C}},
    year = {1991},
    journal = {Measurement Science and Technology},
    author = {Yilbas, B. S. and Danisman, K. and Yilbas, Z.},
    number = {7},
    pages = {668--674},
    volume = {2},
    doi = {10.1088/0957-0233/2/7/016},
    issn = {09570233}
}

@article{Boley2016,
    title = {{Metal powder absorptivity: modeling and experiment}},
    year = {2016},
    journal = {Applied Optics},
    author = {Boley, C. D. and Mitchell, S. C. and Rubenchik, A. M. and Wu, S. S. Q.},
    number = {23},
    pages = {6496},
    volume = {55},
    doi = {10.1364/ao.55.006496},
    issn = {0003-6935}
}

@article{Qin2017,
    title = {{Near-Infrared Plasmonic Copper Nanocups Fabricated by Template-Assisted Magnetron Sputtering}},
    year = {2017},
    journal = {ACS Photonics},
    author = {Qin, Yunxiang and Kong, Xiang Tian and Wang, Zhiming and Govorov, Alexander O. and Kortshagen, Uwe R.},
    number = {11},
    pages = {2881--2890},
    volume = {4},
    doi = {10.1021/acsphotonics.7b00866},
    issn = {23304022}
}

@article{Ikeshoji2018,
    title = {{Selective Laser Melting of Pure Copper}},
    year = {2018},
    journal = {Jom},
    author = {Ikeshoji, Toshi Taka and Nakamura, Kazuya and Yonehara, Makiko and Imai, Ken and Kyogoku, Hideki},
    number = {3},
    pages = {396--400},
    volume = {70},
    publisher = {Springer US},
    url = {https://doi.org/10.1007/s11837-017-2695-x},
    doi = {10.1007/s11837-017-2695-x},
    issn = {15431851}
}

@article{Colopi2018,
    title = {{Selective laser melting of pure Cu with a 1 kW single mode fiber laser}},
    year = {2018},
    journal = {Procedia CIRP},
    author = {Colopi, M. and Caprio, L. and Demir, A.G. and Previtali, B.},
    pages = {59--63},
    volume = {74},
    publisher = {Elsevier B.V.},
    url = {https://linkinghub.elsevier.com/retrieve/pii/S221282711830814X},
    doi = {10.1016/j.procir.2018.08.030},
    issn = {22128271}
}

@article{Jadhav2020,
    title = {{Surface Modified Copper Alloy Powder for Reliable Laser-based Additive Manufacturing}},
    year = {2020},
    journal = {Additive Manufacturing},
    author = {Jadhav, Suraj Dinkar and Dhekne, Pushkar Prakash and Dadbakhsh, Sasan and Kruth, Jean Pierre and Van Humbeeck, Jan and Vanmeensel, Kim},
    number = {June},
    pages = {101418},
    volume = {35},
    publisher = {Elsevier},
    url = {https://doi.org/10.1016/j.addma.2020.101418},
    doi = {10.1016/j.addma.2020.101418},
    issn = {22148604}
}

@article{Baffou2013,
    title = {{Thermo-plasmonics: Using metallic nanostructures as nano-sources of heat}},
    year = {2013},
    journal = {Laser and Photonics Reviews},
    author = {Baffou, Guillaume and Quidant, Romain},
    number = {2},
    pages = {171--187},
    volume = {7},
    doi = {10.1002/lpor.201200003},
    issn = {18638880}
}

@article{Vukusic2020-Nature,
    title = {{- Photonic Structures in Nature}},
    year = {2020},
    journal = {Bionanotechnology II},
    author = {Vukusic, Pete and Sambles, J Roy},
    number = {August},
    pages = {516--537},
    volume = {424},
    doi = {10.1201/b11374-29}
}

@article{Talignani2022AAlloys,
    title = {{A review on additive manufacturing of refractory tungsten and tungsten alloys}},
    year = {2022},
    journal = {Additive Manufacturing},
    author = {Talignani, Alberico and Seede, Raiyan and Whitt, Austin and Zheng, Shiqi and Ye, Jianchao and Karaman, Ibrahim and Kirka, Michael M. and Katoh, Yutai and Wang, Y. Morris},
    number = {June},
    pages = {103009},
    volume = {58},
    publisher = {Elsevier B.V.},
    url = {https://doi.org/10.1016/j.addma.2022.103009},
    doi = {10.1016/j.addma.2022.103009},
    issn = {22148604}
}

@article{Dorow-Gerspach2021AdditiveMelting,
    title = {{Additive manufacturing of high density pure tungsten by electron beam melting}},
    year = {2021},
    journal = {Nuclear Materials and Energy},
    author = {Dorow-Gerspach, D. and Kirchner, A. and Loewenhoff, Th and Pintsuk, G. and Wei{\ss}g{\"{a}}rber, T. and Wirtz, M.},
    number = {February},
    volume = {28},
    doi = {10.1016/j.nme.2021.101046},
    issn = {23521791}
}

@article{Muller2019Additive1000C,
    title = {{Additive manufacturing of pure tungsten by means of selective laser beam melting with substrate preheating temperatures up to 1000C}},
    year = {2019},
    journal = {Nuclear Materials and Energy},
    author = {M{\"{u}}ller, A. V. and Schlick, G. and Neu, R. and Anst{\"{a}}tt, C. and Klimkait, T. and Lee, J. and Pascher, B. and Schmitt, M. and Seidel, C.},
    number = {February},
    pages = {184--188},
    volume = {19},
    publisher = {Elsevier},
    url = {https://doi.org/10.1016/j.nme.2019.02.034},
    doi = {10.1016/j.nme.2019.02.034},
    issn = {23521791}
}

@article{BurresiBright-WhiteLight,
    title = {{Bright-White Beetle Scales Optimise Multiple Scattering of Light}},
    author = {Burresi, Matteo and Cortese, Lorenzo and Pattelli, Lorenzo and Kolle, Mathias and Vukusic, Peter and Wiersma, Diederik S and Steiner, Ullrich and Vignolini, Silvia},
    pages = {1--8},
    doi = {10.1038/srep06075}
}

@article{Khairallah2020ControllingPrinting,
    title = {{Controlling interdependent meso-nanosecond dynamics and defect generation in metal 3D printing}},
    year = {2020},
    journal = {Science},
    author = {Khairallah, Saad A. and Martin, Aiden A. and Lee, Jonathan R. I. and Guss, Gabe and Calta, Nicholas P. and Hammons, Joshua A. and Nielsen, Michael H. and Chaput, Kevin and Schwalbach, Edwin and Shah, Megna N. and Chapman, Michael G. and Willey, Trevor M. and Rubenchik, Alexander M. and Anderson, Andrew T. and Wang, Y. Morris and Matthews, Manyalibo J. and King, Wayne E.},
    number = {6491},
    month = {5},
    volume = {368},
    doi = {10.1126/science.aay7830},
    issn = {0036-8075}
}

@article{Wang2017DenseMelting,
    title = {{Dense Pure Tungsten Fabricated by Selective Laser Melting}},
    year = {2017},
    journal = {Applied Sciences},
    author = {Wang, Dianzheng and Yu, Chenfan and Zhou, Xin and Ma, Jing and Liu, Wei and Shen, Zhijian},
    number = {4},
    month = {4},
    pages = {430},
    volume = {7},
    url = {http://www.mdpi.com/2076-3417/7/4/430},
    doi = {10.3390/app7040430},
    issn = {2076-3417}
}

@article{Young1961EtchCopper,
    title = {{Etch pits at dislocations in copper}},
    year = {1961},
    journal = {Journal of Applied Physics},
    author = {Young, F. W.},
    number = {2},
    pages = {192--201},
    volume = {32},
    doi = {10.1063/1.1735977},
    issn = {00218979}
}

@article{Martin2020GrainAluminum,
    title = {{Grain refinement mechanisms in additively manufactured nano-functionalized aluminum}},
    year = {2020},
    journal = {Acta Materialia},
    author = {Martin, J. Hunter and Yahata, Brennan and Mayer, Justin and Mone, Robert and Stonkevitch, Ekaterina and Miller, Julie and O'Masta, Mark R. and Schaedler, Tobias and Hundley, Jacob and Callahan, Patrick and Pollock, Tresa},
    pages = {1022--1037},
    volume = {200},
    publisher = {Elsevier Ltd},
    url = {https://doi.org/10.1016/j.actamat.2020.09.043},
    doi = {10.1016/j.actamat.2020.09.043},
    issn = {13596454}
}

@article{Qu2021High-precisionCopper,
    title = {{High-precision laser powder bed fusion processing of pure copper}},
    year = {2021},
    journal = {Additive Manufacturing},
    author = {Qu, Shuo and Ding, Junhao and Fu, Jin and Fu, Mingwang and Zhang, Baicheng and Song, Xu},
    number = {PA},
    pages = {102417},
    volume = {48},
    publisher = {Elsevier B.V.},
    url = {https://doi.org/10.1016/j.addma.2021.102417},
    doi = {10.1016/j.addma.2021.102417},
    issn = {22148604}
}

@article{Kurnsteiner2020High-strengthManufacturing,
    title = {{High-strength Damascus steel by additive manufacturing}},
    year = {2020},
    journal = {Nature},
    author = {K{\"{u}}rnsteiner, Philipp and Wilms, Markus Benjamin and Weisheit, Andreas and Gault, Baptiste and J{\"{a}}gle, Eric Aimé and Raabe, Dierk},
    number = {7813},
    pages = {515--519},
    volume = {582},
    publisher = {Springer US},
    url = {http://dx.doi.org/10.1038/s41586-020-2409-3},
    doi = {10.1038/s41586-020-2409-3},
    issn = {14764687},
    pmid = {32581379}
}

@article{Jadhav2020HighlyPowder,
    title = {{Highly conductive and strong CuSn0.3 alloy processed via laser powder bed fusion starting from a tin-coated copper powder}},
    year = {2020},
    journal = {Additive Manufacturing},
    author = {Jadhav, Suraj Dinkar and Fu, Dongmei and Deprez, Maxim and Ramharter, Kristof and Willems, Denise and Van Hooreweder, Brecht and Vanmeensel, Kim},
    number = {August},
    pages = {101607},
    volume = {36},
    publisher = {Elsevier B.V.},
    url = {https://doi.org/10.1016/j.addma.2020.101607},
    doi = {10.1016/j.addma.2020.101607},
    issn = {22148604}
}

@article{Shi2015KeepingAnts,
    title = {{Keeping cool: Enhanced optical reflection and radiative heat dissipation in Saharan silver ants}},
    year = {2015},
    journal = {Science},
    author = {Shi, Norman Nan and Tsai, Cheng Chia and Camino, Fernando and Bernard, Gary D. and Yu, Nanfang and Wehner, Rüdiger},
    number = {6245},
    pages = {298--301},
    volume = {349},
    doi = {10.1126/science.aab3564},
    issn = {10959203},
    pmid = {26089358}
}

@article{Colopi2019LimitsLaser,
    title = {{Limits and solutions in processing pure Cu via selective laser melting using a high-power single-mode fiber laser}},
    year = {2019},
    journal = {International Journal of Advanced Manufacturing Technology},
    author = {Colopi, Matteo and Demir, Ali Gökhan and Caprio, Leonardo and Previtali, Barbara},
    number = {5-8},
    pages = {2473--2486},
    volume = {104},
    isbn = {0017001904015},
    doi = {10.1007/s00170-019-04015-3},
    issn = {14333015}
}

@article{Tertuliano2022Nanoparticle-enhancedFusion,
    title = {{Nanoparticle-enhanced absorptivity of copper during laser powder bed fusion}},
    year = {2022},
    journal = {Additive Manufacturing},
    author = {Tertuliano, Ottman A. and DePond, Philip J. and Doan, David and Matthews, Manyalibo J and Gu, X Wendy and Cai, Wei and Lew, Adrian J},
    month = {3},
    pages = {102562},
    volume = {51},
    publisher = {Elsevier},
    url = {https://doi.org/10.1016/j.addma.2021.102562 https://linkinghub.elsevier.com/retrieve/pii/S2214860421007090},
    doi = {10.1016/j.addma.2021.102562},
    issn = {22148604}
}

@article{Tertuliano2021Nanoparticle-enhancedFusion,
    title = {{Nanoparticle-enhanced Absorptivity of Copper During Laser Powder Bed Fusion}},
    year = {2021},
    journal = {Additive Manufacturing},
    author = {Tertuliano, Ottman A. and DePond, Philip J. and Doan, David and Matthews, Manyalibo J. and Gu, X. Wendy and Cai, Wei and Lew, Adrian J.},
    number = {September 2021},
    pages = {102562},
    volume = {51},
    publisher = {Elsevier B.V.},
    url = {https://doi.org/10.1016/j.addma.2021.102562},
    doi = {10.1016/j.addma.2021.102562},
    issn = {22148604}
}

@article{Teperik2008OmnidirectionalSurfaces,
    title = {{Omnidirectional absorption in nanostructured metal surfaces}},
    year = {2008},
    journal = {Nature Photonics},
    author = {Teperik, T. V. and Garc{\'{i}}a De Abajo, F. J. and Borisov, A. G. and Abdelsalam, M. and Bartlett, P. N. and Sugawara, Y. and Baumberg, J. J.},
    number = {5},
    pages = {299--301},
    volume = {2},
    doi = {10.1038/nphoton.2008.76},
    issn = {17494885}
}

@article{Polman2012PhotonicPhotovoltaics,
    title = {{Photonic design principles for ultrahigh-efficiency photovoltaics}},
    year = {2012},
    journal = {Nature Materials},
    author = {Polman, Albert and Atwater, Harry A.},
    number = {3},
    pages = {174--177},
    volume = {11},
    publisher = {Nature Publishing Group},
    url = {http://dx.doi.org/10.1038/nmat3263},
    doi = {10.1038/nmat3263},
    issn = {14761122},
    pmid = {22349847}
}

@article{Sndergaard2012PlasmonicGrooves,
    title = {{Plasmonic black gold by adiabatic nanofocusing and absorption of light in ultra-sharp convex grooves}},
    year = {2012},
    journal = {Nature Communications},
    author = {S{\o}ndergaard, Thomas and Novikov, Sergey M. and Holmgaard, Tobias and Eriksen, René L. and Beermann, Jonas and Han, Zhanghua and Pedersen, Kjeld and Bozhevolnyi, Sergey I.},
    pages = {1--6},
    volume = {3},
    doi = {10.1038/ncomms1976},
    issn = {20411723}
}

@article{Kravets2008PlasmonicCoatings,
    title = {{Plasmonic blackbody: Almost complete absorption of light in nanostructured metallic coatings}},
    year = {2008},
    journal = {Physical Review B - Condensed Matter and Materials Physics},
    author = {Kravets, V. G. and Schedin, F. and Grigorenko, A. N.},
    number = {20},
    pages = {97--99},
    volume = {78},
    doi = {10.1103/PhysRevB.78.205405},
    issn = {10980121}
}

@article{Vock2019PowdersReview,
    title = {{Powders for powder bed fusion: a review}},
    year = {2019},
    journal = {Progress in Additive Manufacturing},
    author = {Vock, Silvia and Kl{\"{o}}den, Burghardt and Kirchner, Alexander and Wei{\ss}g{\"{a}}rber, Thomas and Kieback, Bernd},
    number = {4},
    pages = {383--397},
    volume = {4},
    publisher = {Springer International Publishing},
    url = {http://dx.doi.org/10.1007/s40964-019-00078-6},
    isbn = {4096401900},
    doi = {10.1007/s40964-019-00078-6},
    issn = {23639520}
}

@article{Hu2020PureMechanism,
    title = {{Pure tungsten and oxide dispersion strengthened tungsten manufactured by selective laser melting: Microstructure and cracking mechanism}},
    year = {2020},
    journal = {Additive Manufacturing},
    author = {Hu, Zhangping and Zhao, Yanan and Guan, Kai and Wang, Zumin and Ma, Zongqing},
    number = {July},
    pages = {101579},
    volume = {36},
    publisher = {Elsevier B.V.},
    url = {http://dx.doi.org/10.1016/j.addma.2020.101579},
    doi = {10.1016/j.addma.2020.101579},
    issn = {22148604}
}

@article{Morcos2022Review:Alloys,
    title = {{Review: additive manufacturing of pure tungsten and tungsten-based alloys}},
    year = {2022},
    journal = {Journal of Materials Science},
    author = {Morcos, Peter and Elwany, Alaa and Karaman, Ibrahim and Arr{\'{o}}yave, Raymundo},
    number = {21},
    pages = {9769--9806},
    volume = {57},
    doi = {10.1007/s10853-022-07183-y},
    issn = {15734803}
}

@article{Guo2019SelectiveProperties,
    title = {{Selective laser melting additive manufacturing of pure tungsten: Role of volumetric energy density on densification, microstructure and mechanical properties}},
    year = {2019},
    journal = {International Journal of Refractory Metals and Hard Materials},
    author = {Guo, Meng and Gu, Dongdong and Xi, Lixia and Zhang, Hongmei and Zhang, Jiayao and Yang, Jiankai and Wang, Rui},
    number = {March},
    pages = {105025},
    volume = {84},
    publisher = {Elsevier},
    url = {https://doi.org/10.1016/j.ijrmhm.2019.105025},
    doi = {10.1016/j.ijrmhm.2019.105025},
    issn = {22133917}
}

@article{Xue2021SelectiveMicrocracks,
    title = {{Selective laser melting additive manufacturing of tungsten with niobium alloying: Microstructure and suppression mechanism of microcracks}},
    year = {2021},
    journal = {Journal of Alloys and Compounds},
    author = {Xue, Juanqin and Feng, Zheng and Tang, Jingang and Tang, Changbin and Zhao, Zhuang},
    pages = {159879},
    volume = {874},
    publisher = {Elsevier},
    url = {https://doi.org/10.1016/j.jallcom.2021.159879},
    doi = {10.1016/j.jallcom.2021.159879},
    issn = {09258388}
}

@article{Xiong2020SelectiveTungsten,
    title = {{Selective Laser Melting and Remelting of Pure Tungsten}},
    year = {2020},
    journal = {Advanced Engineering Materials},
    author = {Xiong, Zhengang and Zhang, Panpan and Tan, Chaolin and Dong, Dongdong and Ma, Wenyou and Yu, Kun},
    number = {3},
    pages = {1--9},
    volume = {22},
    doi = {10.1002/adem.201901352},
    issn = {15272648}
}

@article{Tan2018SelectiveProperties,
    title = {{Selective laser melting of high-performance pure tungsten: parameter design, densification behavior and mechanical properties}},
    year = {2018},
    journal = {Science and Technology of Advanced Materials},
    author = {Tan, Chaolin and Zhou, Kesong and Ma, Wenyou and Attard, Bonnie and Zhang, Panpan and Kuang, Tongchun},
    number = {1},
    pages = {370--380},
    volume = {19},
    publisher = {Taylor {\&} Francis},
    url = {http://doi.org/10.1080/14686996.2018.1455154},
    doi = {10.1080/14686996.2018.1455154},
    issn = {18785514}
}

@inproceedings{Fu2022TheInnovation,
    title = {{The best kept secret in laser additive manufacturing: green lasers, a unique innovation}},
    year = {2022},
    booktitle = {Laser 3D Manufacturing IX},
    author = {Fu, Eliana and Spiegelhalder, Roland and Vogt, Sabrina and Goebel, Marco},
    editor = {Helvajian, Henry and Gu, Bo and Chen, Hongqiang},
    number = {March},
    month = {3},
    pages = {5},
    volume = {1199202},
    publisher = {SPIE},
    url = {https://www.spiedigitallibrary.org/conference-proceedings-of-spie/11992/2614548/The-best-kept-secret-in-laser-additive-manufacturing--green/10.1117/12.2614548.full},
    isbn = {9781510648555},
    doi = {10.1117/12.2614548},
    issn = {1996756X}
}

@article{Rebesan2021TungstenFusion,
    title = {{Tungsten Fabricated by Laser Powder Bed Fusion}},
    year = {2021},
    journal = {BHM Berg- und H{\"{u}}ttenm{\"{a}}nnische Monatshefte},
    author = {Rebesan, Pietro and Bonesso, Massimiliano and Gennari, Claudio and Dima, Razvan and Pepato, Adriano and Vedani, Maurizio},
    number = {5},
    month = {5},
    pages = {263--269},
    volume = {166},
    url = {https://link.springer.com/10.1007/s00501-021-01109-y},
    doi = {10.1007/s00501-021-01109-y},
    issn = {0005-8912}
}

@article{Shimada2016WhatNanostructures,
    title = {{What is the Key Structural Parameter for Infrared Absorption Enhancement on Nanostructures?}},
    year = {2016},
    journal = {Journal of Physical Chemistry C},
    author = {Shimada, Toru and Nakashima, Hiroshi and Kumagai, Yuta and Ishigo, Yuta and Tsushima, Masamichi and Ikari, Akihiko and Suzuki, Yushi},
    number = {1},
    pages = {534--541},
    volume = {120},
    doi = {10.1021/acs.jpcc.5b09315},
    issn = {19327455}
}
\clearpage

\section*{Data Availability}
All core data generated and analysed for this study can be found in this article and its Supplementary Information file. Further source data leading to these core data are available from the corresponding authors without any restrictions. This includes code to process data not essential.

\section*{Acknowledgments}

We thank Thomas Pluschkell, Jeffrey Fingerle, and Mickeal Ades for providing the LLNL copper powders. We acknowledge US patent application 17/706259 of the invention entitled ``Nanotextured metal powders for 3D printing of metals." Part of this work was performed at the Stanford Nano Shared Facilities (SNSF), supported by the National Science Foundation under award ECCS-2026822. Use of the Stanford Synchrotron Radiation Lightsource, SLAC National Accelerator Laboratory, is supported by the U.S. Department of Energy, Office of Science, Office of Basic Energy Sciences under Contract No. DE-AC02-76SF00515. Part of this work was performed under the auspices of the U.S. Department of Energy by Lawrence Livermore National Laboratory under Contract DE-AC52-07NA27344. We acknowledge funding from National Science Foundation through L.C.’s Graduate Research Fellowship Program (NSF GFRP)


\section*{Competing Interest}
We would like to disclose that a patent relevant to this work has been filed (Patent No.:
17706259).

\pagebreak
\section*{FIGURES}
\clearpage

\clearpage

\clearpage

\end{document}